# Microscopy system for in situ sea ice structure and biology observations.

Béatrice Lessard-Hamel[1,2], Marcel Babin[1], Simon Thibault[2]

[1] Takuvik Joint International Laboratory, Université Laval and CNRS, Québec, Canada
[2] Centre d'Optique, Photonique et Laser (COPL) and Département de Physique, de Génie Physique et d'Optique, Université Laval, Québec, Canada

*Correspondence to*: Béatrice Lessard-Hamel (beatrice.lessard-hamel@copl.ulaval.ca)

*Abstract*

Sea ice harbours a rich community of well-adapted microorganisms that inhabit liquid micro-spaces where extreme conditions prevail. Currently at risk under climate change, the sea-ice microbiome holds mysteries about evolution of life on Earth and possibly elsewhere, which require methodological innovation to be unravelled. Gaining microscopic insight into the internal structure and biology of sea ice has traditionally been limited to destructive and extrusive ice core sampling methods. Here we present an in situ microscopic imaging system to observe undisturbed living microorganisms directly within sea the ice matrix. The complex and heterogeneous nature of sea ice, including its water crystal lattice, brine channels, air bubbles, and various impurities, presents engineering challenges for the development of this imaging system. Despite the fragile nature of the sea-ice matrix, we could successfully deploy, test and use the new in situ microscope during a recent expedition on the icepack in Arctic. We collected numerous images of live and intact single-celled and colony-forming diatoms, and documented for the first time at such a high resolution some microphysical features of sea ice. The hardware and software design of the endoscope is presented along with acquisition results of the microstructure and diatom images. These findings collectively demonstrate the potential for this new in situ microscopic imaging system to transform the way we study sea ice and to allow a deeper understanding of its complex microstructure and living microorganisms.

*Introduction*

Sea ice, which covers 14% of the ocean surface, harbors a rich and unique microbiota that may have played an important role in the evolution of life on Earth and may provide clues to the possible presence of life in other icy ocean worlds (Babin et al. 2025). Today, the sea ice microbiota is a key component of polar marine ecosystems that is at risk in the current context of global change.

At the millimeter to micron scale, sea ice is a complex heterogeneous matrix composed of ice crystals, liquid brines, gas bubbles, and particles. The size, shape, and arrangement of these components vary depending on factors such as temperature, salinity, and the growth conditions of the ice (Hunke et al. 2011; Light et al. 2003). Sympagic (in ice) microorganisms live within the sea ice internal network of microspaces filled with liquid brines. The 2D and 3D description of this labyrinth on a scale relevant to microbes is the ultimate frontier in our knowledge of this major Earth biome. The solid, fast changing and heterogeneous nature of sea ice, coupled with its inherent fragility, however present an engineering challenge for imaging the interior of this medium with limited alteration of the microphysical structure and its inhabitants, bacteria, archaea, protists and viruses.

Traditionally, gaining microscopic insight into sea ice and its microbiome has relied on destructive ice core sampling methods, which is prone to thermodynamic disturbances, mechanical damages, losses of solutes and living matter (Notz et al. 2005), cell mortality, and alterations in cellular metabolic processes during sample handling and storage (Lund-Hansen et al. 2020). Optical microscopy was nevertheless used on extracted sea ice to gain new knowledge on microstructures and microbial organisms found in there. With thin sections of both natural and laboratory-produced ice, measurements of brine channel inclusions in terms of size and number (Perovich and Gow 1991; Eicken 1993; Perovich and Gow. 1996; Light et al. 2003) were achieved, and microbes in brine inclusions observed (Krembs et al. 2002). A more recent approach involved the use of X-ray mCT scanning to obtain the first 3D measurement on ice cores of the larger brine channel network and its connectivity (Golden et al. 2007; Obbard et al. 2009; Pringle et al. 2009, Crabeck et al. 2016).

In the present study, we developed an in situ microscope to image the internal microstructures of sea ice and microalgae present in liquid inclusions, particularly diatoms, with minimal disturbance. The system is submersible, and the depth position can be adjusted to obtain a comprehensive vertical profile in sea ice. The illumination was designed to be easily controlled with four modes, providing different contrasts. A user-friendly custom software enables image captures at the micron scale within the sea ice environment in real time.

*Materials and Procedures*

**System overview**
The layout of the system is shown in Figure 1. The optical components are encased in a cylinder designed to perfectly fit in a Mark II ice-core hole of 10cm in diameter. Given the focus of our research on microalgae, most of our observations were made in the bottom 5 cm layer of sea ice. **Figure 1.C** shows an example of an ice core sample and its 5-cm bottom pigmented layer due to the high concentration of sea ice algae which was sampled and examined with Lugol iodine staining and transmission microscopy. Once the Olympus 20x water immersion objective tip of our in situ microscope is brought into contact with the sea ice, it collects light at focus at a working distance of 3.5 mm into the ice with a depth of field of 1.47mm (**Figure 1.B**). Subsequently, a right-angle kinematic mirror redirects the collected plane light upwards, optimizing the setup for a more compact configuration. Finally, the tube lens is an Edmund Optics 175-mm $MgF_2$ achromatic doublet lens that focuses the light onto the BlackflyUSB-3 color 2.3M pixel camera. The illumination system is a Adafruit 12 RGB ring LED with integrated drivers, located around the microscope lens.

Four different illumination types can be achieved including oblique to achieve oblique back-scatter microscopy (OBM), a technique first used by Ford and collaborators (Ford et al. 2012) to capture phase gradient images in tick scattering tissues. In this approach, the intrinsic scattering properties of the sample are exploited to collect backscattered light from different illumination angles, enabling the decoupling of phase and absorption contrasts to reveal both structural and refractive index variations within the sample. With this illumination mode, structures like bubble inclusions showing a large change of refractive index can be imaged. The LEDs are controlled with an Arduino Nano that is fixed inside the microscope cylinder. Two wires for the camera and the Arduino pass through the pole and are connected to a computer **Figure 1.A**. A user interface was developed to control the camera setting and the illumination modes, allowing live imaging. The full imaging system has a field of view of 563mm x 352mm.

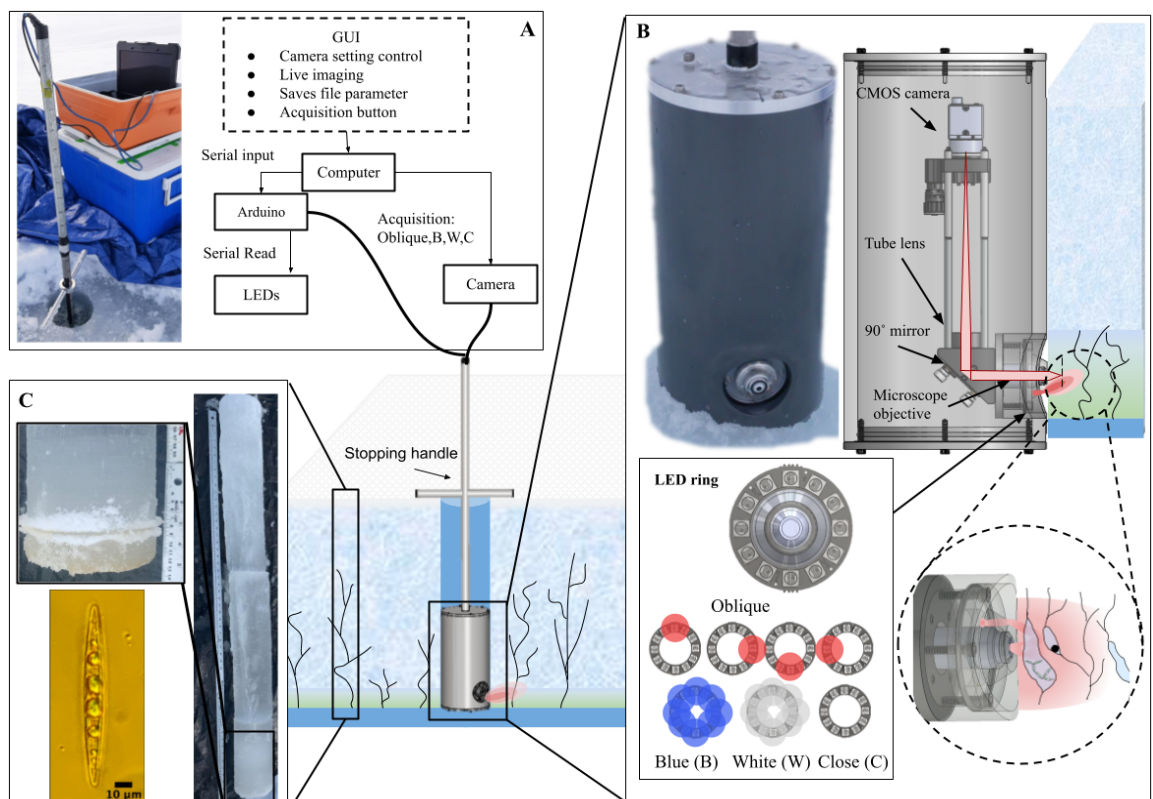

*Figure 1 **The design of the in situ sea ice microscope. The image sampling is conducted horizontally. A. Scheme of the graphical user interface (GUI) to control the illumination and camera acquisition. B. Optical path of the microscope. The objective collects the diffused light for various illumination modes at a working distance of 3.5 mm, imaging brine pockets, air bubbles, and diatoms. C. Ice core sample and transmission microscopy observation of a stained sample from the bottom layer.***

**Mechanical design**
To build the in situ sea ice microscope a mechanical design was performed using the SolidWorks software (2019, Dassault Systems). Since all the optics parts are commercial and well documented, it was straightforward to make the assembly with the supplied .step file of the objective from Edmund optics and all the other parts from Thorlabs .sldprt files. For the illumination system, the LED ring drawing was found on GraCAD (Kyle 2014).

Waterproofing of the enclosure involved the utilization of various-sized rubber O-rings. Specifically, one O-ring was positioned around the water immersion objective tip, while another was placed around the interface between the window and the enclosure to ensure a watertight seal. In addition, two O-rings were applied to both aluminum caps, and Teflon tape was employed for the pole. A Unit Pak 2 unit silica gel bag is installed in the enclosure for humidity control. Previous tests with a vacuum pump were performed for water-leak testing, simulating immersion in depths of up to 5 meters.

**Software**
A Graphical User Interface(GUI) was created in Python using the tkinter, the FLIR python spinnaker and the module simple_pyspin by the Kleckner Lab at UC Merced (Kleckner 2019). This GUI allows controlling at the same time the LED and the Camera during the acquisition. The interface is shown in Supplementary Figure 1 with the default parameters. The camera's pixel format, frame rate, gain, exposure and binning can be adjusted. The objective is 20x by default, but if a 10x is used, the scale bar in the live view automatically changes accordingly. To organize every saved acquisition and keep the file information, the station, site and depth can be changed. The different illumination types can be selected and are automatically changed accordingly. A delay can be added before starting the acquisition up to 10 seconds if other manipulations are needed. The live view window is enabled when the "Start live" button is clicked and shows the scale bar. A console window shows when changes are made and gives information on the acquisitions. All the pictures are saved in .tif files in specific folders depending on the station and site.

*Assessment*

**Field deployment and data collection**
Using the sea ice in situ microscope, we sampled first-year (FY) sea ice at 8 study stations around Qikiqtarjuaq Island next to Baffin Bay in Nunavut, Canada from April 20$^{th}$ and May 3$^{rd}$, 2023 (Supplementary Figure 2). For every station, one or two sites were sampled. We were at the very beginning of the melt season. All the sites were on snow-covered ice ranging from 95 to 143.5 cm and a snow depth ranging from 3 to 10 cm. Air temperature ranged between -11 and 0°C and the sky was sunny with passing clouds for most of the sampling period. All sites had positive freeboard ranging from 2 to 12.5 cm. Sea ice temperature ranged from -5.8 to -1.6˙C with bulk salinity from 3.97 to 10.21 PSU. At the bottom layer of the ice was a visible algae layer. No other impurities were observed by eye in the ice column. Site 8 was particularly interesting being in a ridge area with high spatial variability in ice and snow thickness.

For every site, image acquisitions started in the bottom 10 cm layer at the ice-water interface. For a single depth, acquisitions were made at various directions into the ice. A full profile was achieved with image acquisition every 10 cm. Between approximately 100 to 1000 images were taken at every site for a total of 7 734 images for the 8 stations. A 590 nm long-pass filter (Thorlabs 5 mm OG590 Colored Glass Filter) was inserted for the two first sites but then taken out for a better signal-to-noise ratio.

At every site, ice cores were taken out with a 5.5 inch Ice core *Kovacs* Mark V. Ice thickness and freeboard measurements were made with a Kovacs ice thickness gauge. Ice core temperature was measured at every 10 cm with a temperature probe. Snow thickness was measured with an avalanche ruler. First, a picture of the whole ice core was taken, and then we saw the bottom 3 cm of the ice core while being careful to minimize exposure to light and handling contamination. Those sample were later used for standard microscopy for the sake of comparison with the images from the in situ microscope (Figure 1.C). The other parts were cut at every 10 cm for salinity measurement. The samples were put in a 5L Whirl-Pak® Write-On Sterilized Sampling Bags and put in a dark cooler. Later, the samples from the bottom 3cm were melted in a filtered seawater ratio of 3:1 and fixed with a solution of Lugol iodine to be stored and transported to the lab. The fixed samples were imaged using a sedimentation Utermöhl method with a 20x and 40x transmission microscope.

Each raw image was quality controlled and flagged if they were jugged of good quality.  After being sorted, the good images were given a comment for what could be seen (example: bubble train, algae, chain, other). For the OBM acquisitions, a custom software written in Python was made following the routine described previously (Ford 2012). In the field, we surprisingly managed to get multiple series of images in a short time of acquisition of the same object. For those series, a script was used to create gifs with a framerate of 3Hz which are available in supplementary videos.

*Imaging results*

Figure 2 illustrates the first field results obtained with our in situ sea ice microscope on first-year (FY) sea ice. Figure 2.A and C show the full field of view of two different acquisitions. In Figure 2.A, a long chain of *Nitzschia promare* is captured in blue illumination mode. Long diatom chains (>50 mm) that are fragile and prone to breakage during sampling, due to mechanical stress during ice core extraction and handling, are rarely observed as intact. Here, the full chain remains unbroken and folds on itself in the sea ice matrix. Furthermore, there are different diatom species in proximity shown in the tree closed-up views that were acquired at different focus. This result shows the strong potential for this new method for community interaction analysis. In Figure 2.C, the oblique illumination at the focal plane enables phase gradient images of structures with varying refractive indices that do not absorb light, such as lipid droplets in the cell. A similar *Navicula* sp. from a fixed sample was observed in transmission microscopy for comparison. With this new method, the pigments as well as the structure are kept intact for study.

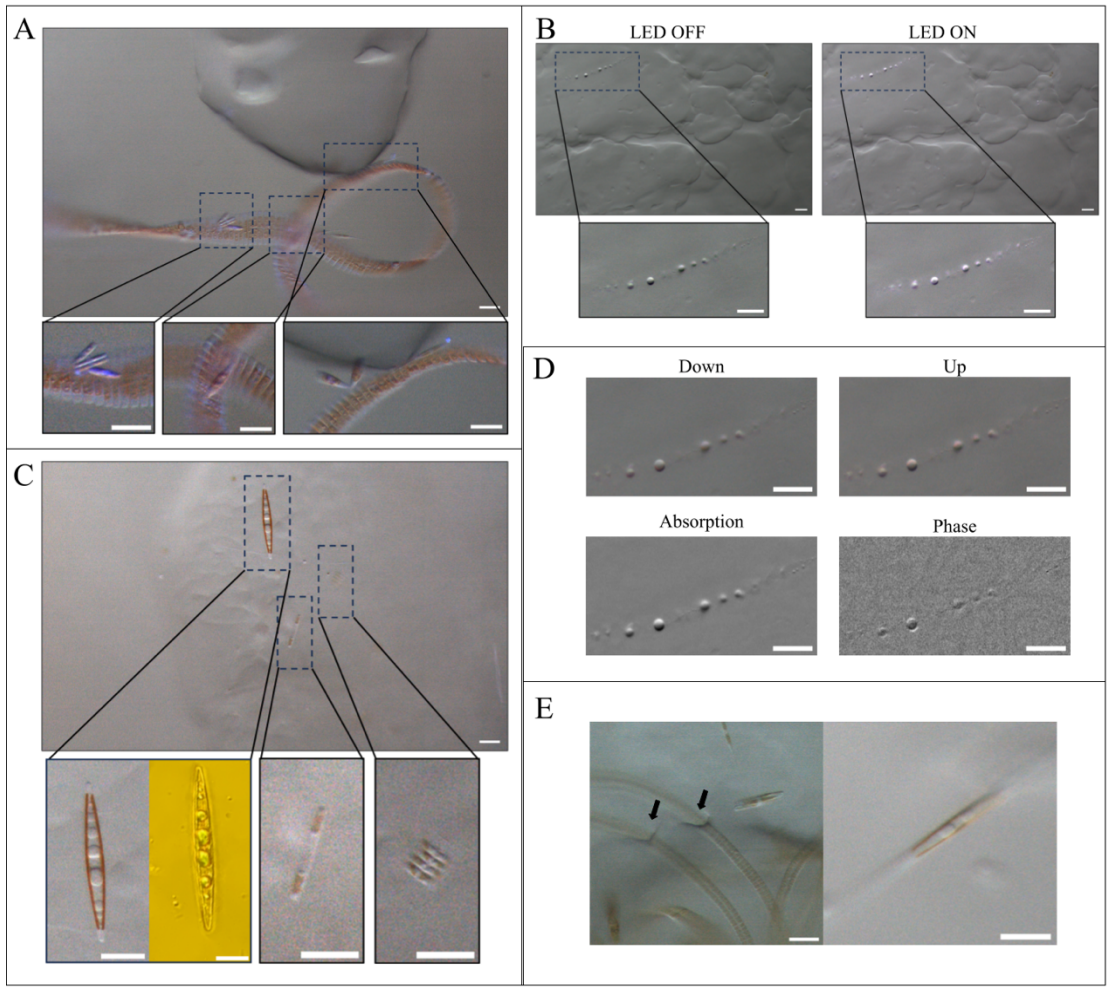

*Figure 2 Microalgae acquisition results of the in situ sea ice microscope. A. Acquisition with blue illumination mode at 100 cm depth in 101.5 cm tick sea ice. Chain of Nitzschia promare with three closed-up from other focal plane showing other species. B. Closed and White illumination at 100 cm depth in 101.5 cm tick sea ice. C. Acquisition with white illumination at 100 depth cm in 102.5 cm thick sea ice. Two closed-up from other focal planes and one close-up of a Navicula sp. compared to a fixed sample. D. Result from oblique illumination mode. Absorption and phase are decoupled from an up-and-down red illumination. E. Examples of algae shape's dependencies with sea ice micro-structures. Arrows pointing to sea ice brine pockets. Scale bar 25mm.*

Because the focal plane of the microscope is located at 3.5 mm away from the lens surface, and because the latter was as much as possible put in contact with sea ice along the sidewall of the coring hole, we believe that the features imaged with our microscope were located within sea ice. One cannot exclude, however, that some of the observed microalgae were instead free floating and inserted between the lens surface and sea ice.

The in situ sea ice microscope was designed not only to image the microorganisms but also the microstructure of the sea ice. In Figure 2.B the ice structure with a bubble train is imaged with off and on white illumination mode. The bubbles have diameters ranging from 1 to 6 mm. From the same field of view with the white illumination mode, smaller bubble trains can now be distinguished from the background. The pulse illumination mode of the microscope, involving capturing images with the LED both on and off, allows low contrast objects like air bubbles to be imaged. In Figure 2.D. the same bubble train is imaged in oblique illumination mode. From the two acquisitions with Down illumination followed by the Up illuminations, the absorption contrast and the phase contrast can be decoupled following the OBM computational process. With this new in situ microscope, brine channels and brine pockets can now be observed without collecting any sample. Figure 2.E shows the potential of the system to now answer questions on the interaction of algae with the ice matrix by looking directly at how the algae are attached to the ice structure. Additional in situ diatom images within their undisturbed environments are available in the data availability section.

The robustness of the in situ sea ice microscope was evaluated from the large number of acquisitions collected under diverse field conditions. Over the eight stations sampled around Qikiqtarjuaq Island, a total of 7,734 images were obtained. After manual sorting and flagging, approximately 12% of the images were considered of sufficient scientific interest for analysis, showing identifiable microstructures or microorganisms. This yield reflects both the natural heterogeneity of the sea ice matrix and the challenge of precise focusing in situ.

Because the microscope must be positioned manually within the ice and the focus adjusted by the operator, image quality depends partly on operator handling. Variations in applied pressure, positioning angle, and subjective judgment during focusing can influence the outcome. Nevertheless, the consistent acquisition of high-quality images across all sites demonstrates that, once the instrument is properly positioned, the method is stable and reproducible enough for systematic field observations when used by trained operators.

*Discussion*

The in situ sea ice microscope provides new opportunities to directly observe the internal microstructures of sea ice and the microorganisms that inhabit them, with limited disruption of the ice matrix. The field results demonstrate that the system can capture intact and living microalgae communities embedded within their natural environment. This capability represents a major improvement over traditional ice core extraction and laboratory imaging methods, which inevitably modify the thermodynamic and biological integrity of samples. The observation of unbroken *Nitzschia promare* chains and intact *Navicula sp.* cells confirms that the in situ configuration preserves delicate organisms and the surrounding brine inclusions, allowing a more accurate representation of community organization and spatial relationships.

Beyond biological observations, the method allows visualization of the physical microstructure of the ice, including bubble trains and brine channels, and their direct association with algal colonies. This offers a unique perspective on how microorganisms occupy and interact with microhabitats, and how the architecture of ice may constrain or facilitate biological activity. The combination of multiple illumination modes, including oblique backscatter, makes it possible to distinguish both phase and absorption features, expanding the range of detectable structures compared to conventional transmitted-light microscopy.

The success of this approach demonstrates that micron-scale, non-destructive imaging of sea ice is feasible under field conditions. However, the efficiency of data collection remains partly dependent on the operator's handling of the instrument. Because positioning and focusing are performed manually within an irregular ice surface, small variations in contact angle or pressure can affect image clarity and framing. Developing standardized acquisition procedures and short training protocols will likely increase the proportion of high-quality images and improve reproducibility among users.

The required resolution for diatom taxonomy classification depends on the level of detail needed to accurately identify and classify the species. To reach a high taxonomic level, ice algae must inevitably be sampled and processed through a sequence of laboratory manipulations, including filtration, staining, and possibly chemical cleaning. Following these preparatory steps, samples are imaged at high resolution by either careful optical microscopy, or electron microscopy yielding nanometer-scale detail for precise taxonomic identification. Such a level of classification is unattainable using an in situ microscope. The development of an in situ sea ice microscope therefore involves a trade-off between a wide field of view and the resolution required for fine-scale taxonomy. As such, our design provides valuable information about algae shape and size, allowing general taxonomic identification such as *Nitzschia, Navicula, Cylindrotheca*, and *Achnanthes*, as demonstrated in the field results.

Despite these limitations, the method bridges an important gap between bulk-scale structural imaging techniques, such as X-ray CT, and traditional microscopy on extracted samples. One of the most spectacular observations we made with our microscope is very long and seemingly intact chains of diatoms, rarely reported before. It also provides a direct observational link between the physical microstructure of the ice and the living communities it supports. This capability opens new avenues for studying microbial ecology, nutrient exchange, and habitat dynamics within sea ice, and for monitoring how these parameters evolve with seasonal and climatic changes. The in situ sea ice microscope thus establishes a foundation for future microscale investigations in the polar cryosphere, where the coupling of physical and biological processes remains a frontier of research.

*Comments and recommendations*

More could be done to improve the in situ sea ice microscope to yield even more valuable insights. Firstly, augmenting the mechanical stability for field acquisition can be achieved by anchoring the microscope stage onto the ice and automating the process of lowering the microscope into the ice core hole. By doing so, one would eliminate user-induced disturbances that might compromise image quality. Secondly, software-controlled focus adjustment could be integrated in the system by mounting the camera on a translation stage with micro-increment movement. In the present system, the picture acquisition is limited to only one focal plane. By motorizing the camera translation, it would become feasible to capture images across various focal planes. This improvement would not only enhance image quality but also facilitate taxonomical identification. Thirdly, a software improvement that could add significant valuable information is a video option. During the field test, previously undocumented microorganism movements and interactions within the sea ice were observed by collecting pictures at high frequency. Introducing video acquisition as an option would significantly enhance the information obtained and open up new possibilities for studying sea ice algae in their environment. This dynamic point of view would allow a more comprehensive understanding of ice algae behavior compared to static images. Collectively, these enhancements would expand the potential of the in situ sea ice microscope.

While the developed imaging system prioritizes non-destructiveness, reducing the impact on the fragile sea ice matrix remains a challenge. The process of ice core drilling introduces potential microstructural alterations and biological impacts. Drilling can create fractures that extend beyond the immediate area of the ice core hole, thereby affecting the surrounding ice and potentially altering its microstructure in a larger region. Moreover, drilling friction can cause localized heating at the hole's edges, slightly melting and then refreezing the ice, which modifies crystal structures and possibly the biological microenvironment. Despite these challenges, the developed system succeeded in providing a close-to-undisturbed visualization at the micron scale of the sea ice microstructure and biological constituents, thereby offering valuable insights.

Another aspect to consider involves the imaging depth achieved within the sea ice, which is dependent on environmental conditions and microscope manipulation. The process of ice coring, particularly when the ice is more than one meter thick and must be cut into segments, can introduce potential alterations to the diameter of the ice hole. Factors such as drilling mechanics, blade condition, ice density, and temperature all contribute to the variability in hole dimensions. Consequently, during image acquisition, gauging the precise depth at which the microscope operates presents a challenge. In addition, slight susceptibility to tilt can lead to small deviations from perfect vertical alignment during imaging. Despite these considerations, the in situ sea ice microscope remains a new scientific tool that offers valuable insights into this complex and delicate environment previously undocumented.

*Acknowledgments*

We would like to thank Guislain Bécu for his contributions to the mechanical design of the microscope and for providing technical support throughout the entire development process. We also thank Marie-Hélène Forget for her valuable logistical support during Arctic field deployment and throughout the project. We acknowledge Daniel Coté for his advice on imaging techniques and Denis Brousseau for his insightful guidance on the OBM optical simulations. We thank Flavienne Bruyant for her logistical assistance during laboratory tests and Joanie Ferland for her assistance during field operations. We also thank Bastian Raulier, Christophe Perron, and Félix Levesque-Desrosiers for the stimulating and insightful discussions. This research project was supported by the SMAART program, funded by the Natural Sciences and Engineering Research Council of Canada, and by the Sentinel North program of Université Laval, made possible in part through funding from the Canada First

Research Excellence Fund. We gratefully acknowledge the scientific and financial support provided by Québec-Océan. Field measurement equipment and related funding were provided through the Polar Knowledge Canada program. Finally, part of this manuscript benefited from the use of DeepL to improve spelling and grammar.

*Data Availability*
Images from station 4 and 8 supporting the findings of this study are available via figshare (https://figshare.com/account/home#/projects/256370 ) and data from all other station are available upon request.

*Code Availability*
A custom interface with serial communication with the Camera and the Arduino is available on Github: https://github.com/BeatriceLH/Insitu-Micro-Sea-Ice-GUI.git

**Supplementary Figures.**

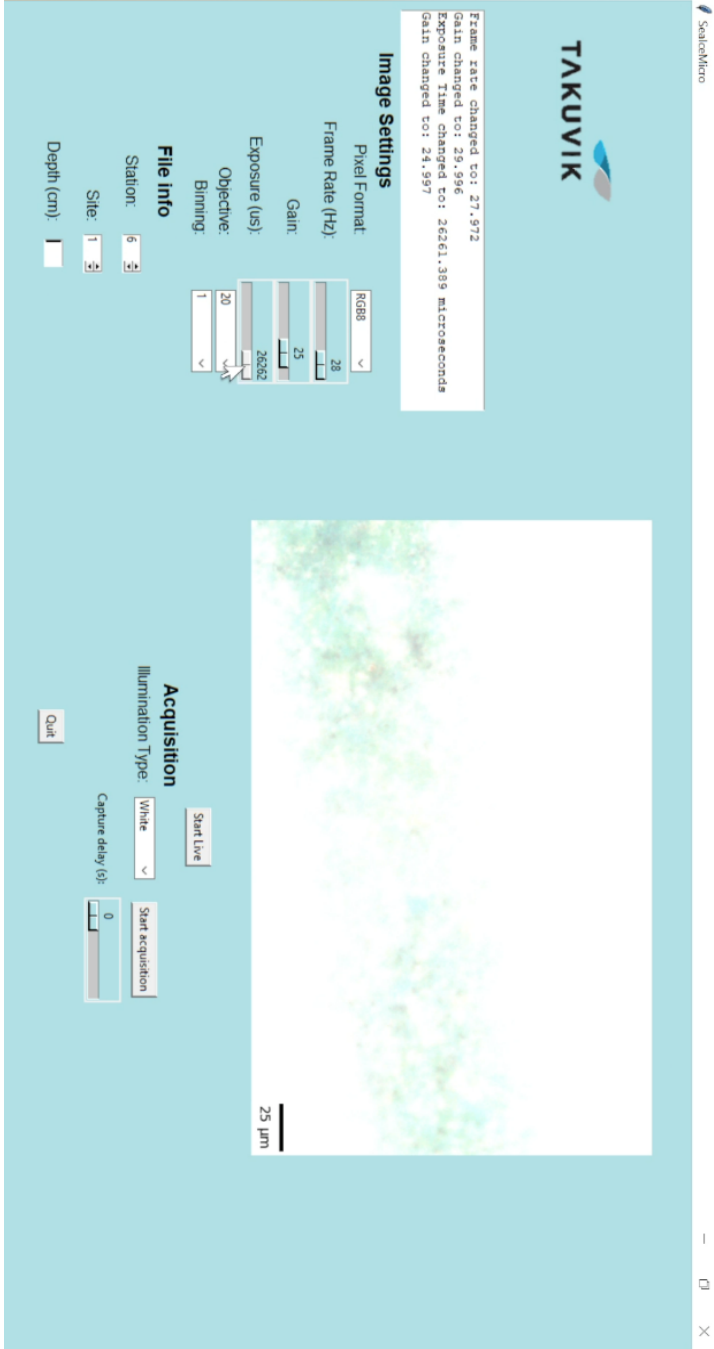

*Supplementary Figure 1 Microscope acquisition GUI*

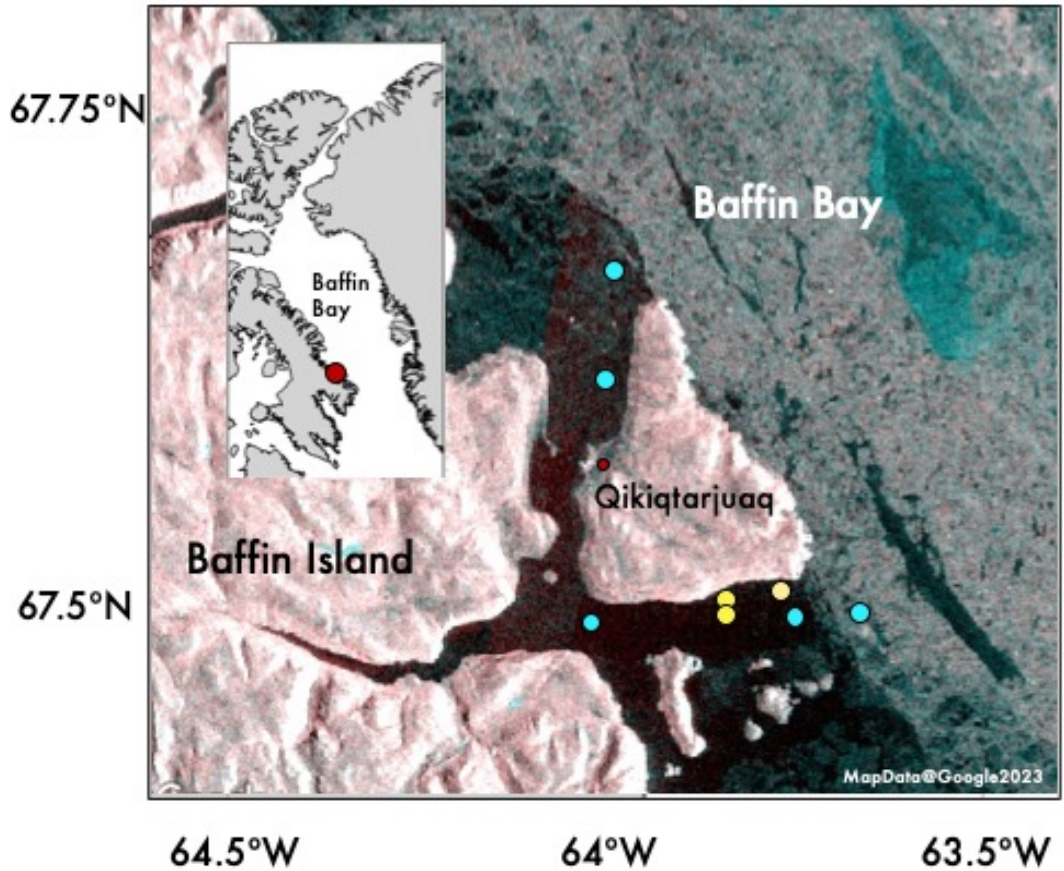

*Supplementary Figure 2. Map of sampled stations[5]. The stations are color-coded according to the depth of the snow layer: yellow indicates a thin layer (less than 10 cm) and blue indicates a thick layer (greater than 10 cm).*